\documentclass[onecolumn,showpacs,preprintnumbers,amsmath,amssymb]{revtex4}
\usepackage{amsmath}
\usepackage{amssymb}
\usepackage{epsfig}
\usepackage{graphicx}
\usepackage{epsfig}
\usepackage{subfigure}
\begin{document}
\def\pslash{\rlap{\hspace{0.02cm}/}{p}}
\def\eslash{\rlap{\hspace{0.02cm}/}{e}}
\title { Top quark rare three-body decays in the
littlest Higgs model with T-parity }
\author{Jinzhong Han}
\author{Bingzhong  Li}
\author{Xuelei Wang}

\affiliation{College of Physics and Information Engineering, Henan
Normal University, Xinxiang, Henan 453007. P.R. China}
\begin{abstract}
In the littlest Higgs model with T-parity (LHT), the mirror quarks
have flavor structures and will contribute to the top quark flavor
changing neutral current. In this work, we perform an extensive
investigation of the top quark rare three-body decays $t\rightarrow
cVV~(V=\gamma,Z,g)$ and $t\rightarrow cf\bar{f}~(f=b,\tau,\mu,e)$ at
one-loop level. Our results show that the branching ratios of
$t\rightarrow cgg$ and $t\rightarrow cb\bar{b}$ could reach
$\mathcal {O}(10^{-3})$ in the favorite parameter space of the
littlest Higgs model with T-parity, which implies that these decays
may be detectable at the LHC or ILC,  while for the other decays,
their rates are too small to be observable at the present or future
colliders.

\end{abstract}


\maketitle


\section{ Introduction}
Top quark physics is among the central physical topics at the
Tevatron and will continue to be so at the Large Hadron Collider
(LHC) in the next few years. Compared to other lighter SM fermions,
the top quark is the only fermion with mass at the electroweak
symmetry breaking scale, so it is widely speculated that the
properties of the top quark are sensitive to new physics. Among
various top quark processes at present and future colliders, the
flavor changing neutral current (FCNC) processes are often utilized
to probe new physics (NP) because in the SM, the FCNC processes are
highly suppressed \cite{1}, while in NP models, there may be no such
suppression. Therefore, searching for top FCNC at colliders can
serve as an effective way to hunt for NP.

The two-body FCNC decays of top quark such as $ t\rightarrow cg,
c\gamma, cZ, cH$ received much attention in the past.  In the SM,
the rates of these decays are less than $10^{-11}$ \cite{sm-2t},
which is far below the reaches of the LHC
\cite{limit-LHC1,limit-LHC2} and the International Linear Collider
(ILC) \cite{limit-ILC}, while in many NP models these decays may be
enhanced to detectable levels \cite{detectable}. By now, the
two-body processes $ t\rightarrow cg, c\gamma, cZ, cH$ have been
extensively investigated in the minimal supersymmetric standard
model (MSSM) \cite{MSSM-2t}, the left-right supersymmetric models
\cite{LR-SUSY}, the supersymmetric model with R-parity violation
\cite{SUSY-RV}, the two-Higgs doublet model (2HDM) \cite{2HDM-2t},
the topcolor-assisted technicolor model (TC2) \cite{TC2-2t}, as well
as models with extra singlet quarks \cite{Extra-quark}. Beside this,
some three-body FCNC decays of the top quark, such as  $t\rightarrow
cVV~(V=\gamma,Z,g)$ and $t\rightarrow cf\bar{f}~(f=b, \tau, \mu,
e)$, were also studied in the framework of the SM \cite{SM-3t,
SM-3t-domin.,SM-2HDM-3t,SM-MSSM-2HDM-3t}, 2HDM
\cite{SM-2HDM-3t,2HDM-3t,SM-MSSM-2HDM-3t}, MSSM
\cite{MSSM-3t,SM-MSSM-2HDM-3t,RPV-MSSM,Eff.-vert.-MSSM-3t}, TC2
\cite{TC2-tcVV,TC2-tcgg,tcll-TC2,tcbb-TC2}, or in a
model-independent way \cite{ind-3t}.

The aim of this work is to perform a comprehensive analysis of the
FCNC  top quark decays $t\rightarrow cVV~(V=\gamma,Z,g)$ and
$t\rightarrow cf\bar{f}~(f=b, \tau, \mu, e)$ in the little Higgs
model with T-parity (LHT) \cite{LHT}. In the LHT model, the related
two-body decays $t\rightarrow cg, c\gamma, cZ, cH$ and the
three-body decays $t\rightarrow cl\bar{l}~(l=\tau,\mu,e)$ have been
studied in \cite{LHT-2t,LHT-3t} respectively, and these studies show
that, compared with the SM, the rates of these decays can be greatly
enhanced. So taking the completeness and the phenomenon of higher
order dominance into consideration \cite{SM-3t-domin.},  it is
necessary to consider all the three-body decays, which will be done
in this work.

\indent  This paper is organized as follows. In Sec II a brief
review of the LHT is given. In Sec III we present the details of our
calculation of the decays $t\rightarrow cVV$ and $t\rightarrow
cf\bar{f}$, and show some numerical results. Finally, we give a
short conclusion in Sec IV.

\section{ A brief review of the LHT}

One of the major motivations for the little Higgs model
\cite{LH1,LH2} is to resolve the little hierarchy problem
\cite{Hierarchy}, in which the quadratic divergence of the Higgs
mass term at one-loop level was canceled by the new diagrams with
additional gauge bosons and a heavy top-quark partner. It was soon
recognized that the scale of the new particles should be in the
multi-TeV range in order to satisfy the constraints from electroweak
precision measurements, which in turn reintroduces the little
hierarchy problem \cite{Reintr.}.  This problem has been eased in
the LHT model where a new $\mathbb{Z}_2$ discrete symmetry called
``T-parity" is introduced, and in this way, all dangerous tree level
contribution to the precision measurements are forbidden \cite{LHT}.

Just like the little Higgs model, in the LHT model the assumed
global symmetry $SU(5)$ is spontaneously broken down to $SO(5)$ at a
scale $f\sim\mathcal {O}(TeV)$, and the embedded $[SU(2)\otimes
U(1)]^2$ gauge symmetry is simultaneously broken at $f$ to the
diagonal subgroup $SU(2)_{L}\otimes U(1)_{Y}$, which is identified
with the SM gauge group. From  the $SU(5)/SO(5)$ breaking, there
arise 14 Goldstone bosons which are described by the ``pion" matrix
$\Pi$, given explicitly by
\begin {equation}
\Pi=
\begin{pmatrix}
-\frac{\omega^0}{2}-\frac{\eta}{\sqrt{20}}&-\frac{\omega^+}{\sqrt{2}}
&-i\frac{\pi^+}{\sqrt{2}}&-i\phi^{++}&-i\frac{\phi^+}{\sqrt{2}}\\
-\frac{\omega^-}{\sqrt{2}}&\frac{\omega^0}{2}-\frac{\eta}{\sqrt{20}}
&\frac{v+h+i\pi^0}{2}&-i\frac{\phi^+}{\sqrt{2}}&\frac{-i\phi^0+\phi^P}{\sqrt{2}}\\
i\frac{\pi^-}{\sqrt{2}}&\frac{v+h-i\pi^0}{2}&\sqrt{4/5}\eta&-i\frac{\pi^+}{\sqrt{2}}&
\frac{v+h+i\pi^0}{2}\\
i\phi^{--}&i\frac{\phi^-}{\sqrt{2}}&i\frac{\pi^-}{\sqrt{2}}&
-\frac{\omega^0}{2}-\frac{\eta}{\sqrt{20}}&-\frac{\omega^-}{\sqrt{2}}\\
i\frac{\phi^-}{\sqrt{2}}&\frac{i\phi^0+\phi^P}{\sqrt{2}}&\frac{v+h-i\pi^0}{2}&-\frac{\omega^+}{\sqrt{2}}&
\frac{\omega^0}{2}-\frac{\eta}{\sqrt{20}}
\end{pmatrix}
\end{equation}
Among the Goldstone bosons,  the fields $\omega^0, \omega^\pm$ and
$\eta$ are eaten by the new heavy gauge bosons $Z_H$, $W^{\pm}_H$
and $A_H$ so that the gauge bosons acquire following masses:
\begin{equation}
M_{W_{H}^{\pm
}}=M_{Z_{H}}=gf(1-\frac{v^2}{8f^2}),~~~~~~~~M_{A_H}=\frac{g'}{\sqrt{5}}f(1-\frac{5v^2}{8f^2}).
\end{equation}
Likewise, the fields $\pi^0$ and $\pi^\pm$ are eaten by the SM gauge
bosons $Z$ and $W^{\pm}$, but one minor difference from the SM is
the masses of these bosons, up to $\mathcal {O}(v^2/f^2)$, are given
by
\begin{equation}
M_{W_{L}}=\frac{gv}{2}(1-\frac{v^2}{12f^2}),~~~~~~~~M_{Z_L}=\frac{gv}{2\cos\theta_W}(1-\frac{5v^2}{12f^2}),
\end{equation}
where $g$ and $g'$ are the SM $SU(2)$ and $U(1)$ gauge couplings
respectively, and $v = 246 {\rm GeV}$.

In the framework of the LHT model, all the SM particles are assigned
to be T-parity even, and the other particles, such as the new gauge
bosons, are assigned to T-parity odd. In particular, in order to
implement the T-parity symmetry, each SM fermion must be accompanied
by one heavy fermion  called the mirror fermion. In the following,
we denote the mirror fermions by $u_{H}^{i}$ and $d_{H}^{i}$ with $i
=1, 2, 3$ being the generation index. At the order of $\mathcal
{O}(v^2/f^2)$, their masses are given by
\begin{eqnarray}
m^i_{d_H}=\sqrt{2}\kappa_if,~~~~~~
m^i_{u_H}=m^i_{d_H}(1-\frac{v^2}{8f^2}),
\end{eqnarray}
where the Yukawa couplings $\kappa_i$ generally depend on the
fermion species $i$.

Since the T-parity is conserved in the LHT model, the fermion pairs
interacting with the T-odd gauge boson must contain one SM fermion
and one mirror fermion. In this case, due to the misalignment of the
mass matrices for the SM fermions and for the mirror fermions, new
gauge bosons can mediate flavor changing interactions. As pointed
out in \cite{FCNC-LHT,Feyn.-rules}, these interactions can be
described by two correlated CKM-like unitary mixing matrices
$V_{H_u}$ and $V_{H_d}$ satisfying
$V_{H_u}^{\dagger}V_{H_d}=V_{CKM}$ with the subscripts $u$ and $d$
denoting which type of the SM fermion is involved in the
interaction. The details of the Feynman rules for such interactions
were given in Ref. \cite{Feyn.-rules}, and in order to clarify our
results, we list some of them:
\begin{eqnarray}
 && \bar{u}^i_H\eta u^j:~-\frac{ig'}{10m_{A_H}}(m^u_{Hi}P_L-m^j_{u}P_R)(V_{H_u})_{ij}, \\
 && \bar{u}^i_H\omega^0u^j:~\frac{ig'}{2m_{Z_H}}(m^u_{Hi}P_L-m^j_{u}P_R)(V_{H_u})_{ij},\\
 && \bar{d}^i_H\omega^-u^j:~\frac{g}{\sqrt{2}m_{W_H}}(m^d_{Hi}P_L-m^j_{u}P_R)(V_{H_u})_{ij},\\
 && \bar{u}^i_HA_H u^j:~-\frac{ig'}{10}(V_{H_u})_{ij}\gamma^\mu P_L,\\
 && \bar{u}^i_HZ_H u^j:~-\frac{ig}{2}(V_{H_u})_{ij}\gamma^\mu P_L,\\
 && \bar{d}^i_HW^{-\mu}_Hu^j:~\frac{ig}{\sqrt{2}}(V_{H_u})_{ij}\gamma^\mu P_L.
\end{eqnarray}

The unitary matrix $V_{H_d}$ is usually parameterized with three
angles $\theta^d_{12},~\theta^d_{23},~\theta^d_{13}$ and three
phases $\delta^d_{12},~\delta^d_{23},~\delta^d_{13}$ \cite{FC-LHT0}:
\begin{eqnarray}
V_{H_d}=
\begin{pmatrix}
c^d_{12}c^d_{13}&s^d_{12}c^d_{13}e^{-i\delta^d_{12}}&s^d_{13}e^{-i\delta^d_{13}}\\
-s^d_{12}c^d_{23}e^{i\delta^d_{12}}-c^d_{12}s^d_{23}s^d_{13}e^{i(\delta^d_{13}-\delta^d_{23})}&
c^d_{12}c^d_{23}-s^d_{12}s^d_{23}s^d_{13}e^{i(\delta^d_{13}-\delta^d_{12}-\delta^d_{23})}&
s^d_{23}c^d_{13}e^{-i\delta^d_{23}}\\
s^d_{12}s^d_{23}e^{i(\delta^d_{12}+\delta^d_{23})}-c^d_{12}c^d_{23}s^d_{13}e^{i\delta^d_{13}}&
-c^d_{12}s^d_{23}e^{i\delta^d_{23}}-s^d_{12}c^d_{23}s^d_{13}e^{i(\delta^d_{13}-\delta^d_{12})}&
c^d_{23}c^d_{13}  \label{mixing}
\end{pmatrix}
\end{eqnarray}
and with the relation $V_{H_u}^{\dagger}V_{H_d}=V_{CKM}$, one can
determine the expression of $V_{H_u}$.

\begin{figure}
\begin{center}
\includegraphics [scale=0.5] {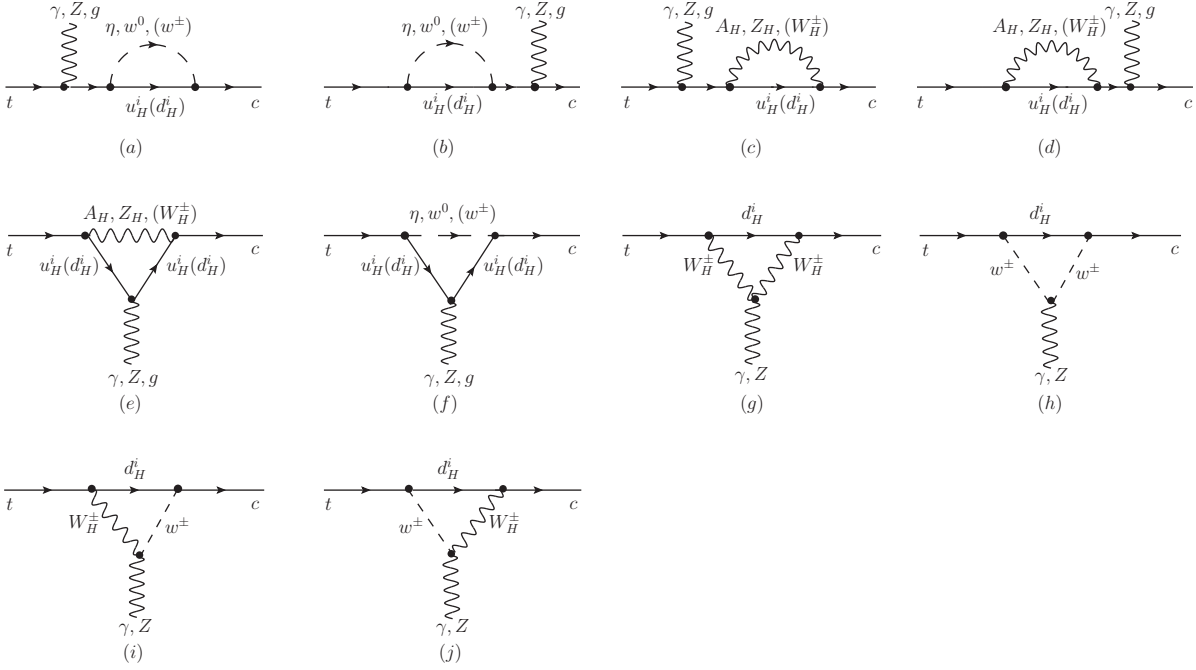}
\caption{The Feynman diagrams of the LHT model contributing to the
FCNC couplings $t\bar{c}V~(V=\gamma,Z,g)$.} \label{fig:fig2}
\end{center}
\end{figure}

\section{Calculations}

\subsection{The loop-level FC couplings $t\bar{c}V~(V=\gamma,Z,g)$
in the LHT model}

As introduced above, in the LHT model new contributions to the FCNC
top quark coupling $t\bar{c}V$ come from the new gauge interactions
mediated by ($A_H,Z_H,W^{\pm}_H$), which are shown in Fig. 1. Since
we use Feynman gauge in our calculation, the Goldstone bosons
$\eta$, $\omega^0$ and $\omega^{\pm}$ also appear in the diagrams.
The heavy scalar triplet $\Phi$, in principle, may also contribute
to the FCNC coupling,  but since such a contribution is suppressed
by the factor ${v^2}/{f^2}$, we neglect it hereafter. It should be
noted that the rules in (5)-(10) imply that the form factors of the
loop-induced $t\bar{c}V$ interaction, $F$, must take the following
form
\begin{eqnarray}
F &\propto& \sum_{i=1}^3 \left ({V^\dag_{H_u}}\right)_{ti} f(m_{Hi})
\left({V_{H_u}}\right)_{ic}  \label{formfactor}
\end{eqnarray}
where $f(m_{Hi})$ is a universal function for three generation
mirror quarks, but its value depends on the mass of $i$th-generation
mirror quark, $m_{Hi}$. Obviously, for the degeneracy of the three
generation mirror quarks, $F$ vanished exactly due to the unitary of
$V_{H_u}$, while for the degeneracy of the first two generations as
discussed below, the factor behaviors like  $( V^\dag_{H_u} )_{t3}
\left ( f(m_{H3}) - f(m_H) \right ) \left({V_{H_u}}\right)_{3c} $
with $m_H$ being the common mass of the first two generations. In
the case of very heavy third generation mirror quarks, $f(m_{H3})$
vanish, that is its effect decouples,  then $F$ is proportional to $
( V^\dag_{H_u} )_{t3} f(m_H) \left({V_{H_u}}\right)_{3c} $, which
are independent of $m_{H3}$.

The Feynman diagrams for the top quark decays $t\rightarrow cVV$ and
$t\rightarrow cf\bar{f}$ are shown in Fig. 2 with the black square
denoting the loop-induced $t\bar{c}V$ vertex. One important
difference of the effective $t\bar{c}V$ verteices in Figs. 2(a, d,
e) from those in Fig. 2(b, c) is for the former cases, both top and
charm quarks are on-shell, while for the latter case,  either top or
charm quark is off-shell. In order to simplify our calculation, we
adopt the calculation method introduced in \cite{Eff.-vert.-MSSM-3t}
which uses a universal form of the effective $t\bar{c}V$ verteices,
but is valid for all the cases. In Appendix A we give the analytical
expressions of the effective verteices $t\bar{c}V$ and use the codes
LoopTools \cite{LoopTools} to get the numerical results of the
relevant loop functions.  To secure the correctness of our results,
we recalculated the two-body decay $t\rightarrow cV$ and find our
results agree with those in Ref. \cite{LHT-2t}.

\begin{figure}
\begin{center}
\includegraphics [scale=0.5] {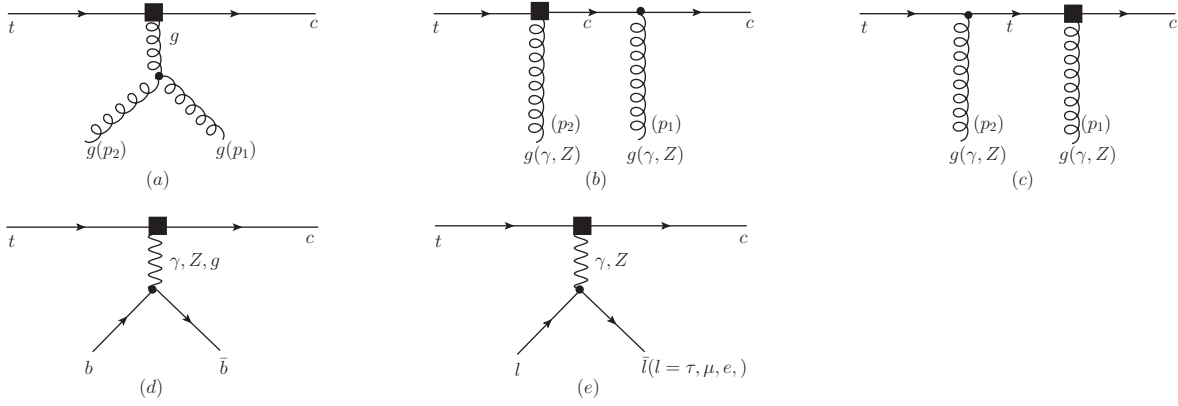}
\caption{The Feynman diagrams for the decays $t\rightarrow cVV$
and $t\rightarrow cf\bar{f}(f=b,\tau,\mu,e)$ in the LHT.}
\label{fig:fig1}
\end{center}
\end{figure}
\subsection{ Amplitude for $t\rightarrow cVV$
in the LHT model}

Since the expressions of the amplitudes for $t\rightarrow cgg,
cg\gamma,cgZ,c\gamma\gamma$  are quite similar, we only list the
result for $t\rightarrow cgg$,  which is given by
 \begin{eqnarray}
\mathcal {M} (t\rightarrow cgg)=\mathcal {M}^g_{a}+\mathcal
{M}^g_{b}+\mathcal {M}^g_{c}
\end{eqnarray}
with
\begin{eqnarray}
\mathcal{M}^g_{a}&=&-ig_sf^{abc}G(p_t-p_c,0)\bar{u}^i_{c}(p_c)\Gamma^{\mu
  cji}_{tcg}
 [(p_1-p_2)_\mu\varepsilon^{a}(p_1)\cdot\varepsilon^{b}(p_2)
 +2p_2\cdot\varepsilon^{a}(p_1)\varepsilon^b_\mu(p_2)\nonumber~~~~\\&&
 -2p_1\cdot\varepsilon^{b}(p_2)\varepsilon^{a}_\mu(p_1)]{u_{t}^j}(p_t)
 \\
\mathcal {M}^g_{b}&=&g_sT^{aki}G(p_t-p_2,m_c)
 \bar{u}^i_{c}(p_c)\rlap/\varepsilon^{a}_1(p_1)(\pslash_t-\pslash_2+m_c)
 \Gamma^{\mu
 bjk}_{tcg}(p_t-p_2,p_c)\varepsilon^{b}_\mu(p_2){u_{t}^j}(p_t) \\
\mathcal {M}^g_{c}&=&g_sT^{bjk}G(p_t-p_2,m_t)
 \bar{u}^i_{c}(p_c)\Gamma^{\mu aki}_{tcg}(p_c,p_t-p_1)
 \varepsilon^{a}_\mu(p_1)(\pslash_t-\pslash_2+m_t)\rlap/\varepsilon^{b}_2(p_2)
 {u_{t}^j}(p_t)~~~~
\end{eqnarray}
In above expressions, $P_{L,R}=\frac{1}{2}(1\mp\gamma_5)$ are the
left and right chirality projectors, $p_t$ is the top quark
momentum, $p_c, p_1, p_2$ are the momentum of the charm quark and
gluons respectively, $\varepsilon$s are wave functions of the
gluons, and  $G(p, m)$ is defined as $\frac{1}{p^2-m^2}$. In actual
calculation, we compute the amplitudes numerically by using the
method of Ref. \cite{Eff.-vert.-MSSM-3t}, instead by calculating the
amplitude square analytically. This greatly simplifies our
calculations.

\subsection{Numerical results for $t\rightarrow cVV$ and $t\rightarrow cf\bar{f}$
in the LHT model}

In this work, we take the SM parameters as: $m_t$ = 172.0 GeV, $m_c$
= 1.27 GeV,  $m_{e}=$0.00051 GeV, $m_{\mu}=$0.106 GeV,
$m_\tau=$1.777 GeV, $m_b$ = 4.2 GeV, $m_Z$ = 91.2 GeV,
$sin^{2}\theta_{W}$ = 0.231, $\alpha_e$=1/128.8,
$\alpha_s(m_t)$=0.107 \cite{SM-paramet.}. For the parameters in the
LHT model, the breaking scale $f$, the three generation mirror quark
masses $m_{H_i}(i=1,2,3)$ and six mixing parameters ($\theta^d_{ij}$
and $\delta^d_{ij}$ with $i,j=1,2,3$ and $i\neq j$)
 in Eq. (\ref{mixing}) are involved.
The breaking scale $f$ determines the new gauge boson masses, and it
has been proven that,  as long as $f\geq500$ GeV, the LHT model can
be consistent with the precision electroweak data\cite{EW
constraint}. So we set $f=500{\rm GeV}, 1000{\rm GeV}$ as two
representative cases. The matrix elements of $V_{H_d} $ have been
severely constrained by the FCNC processes in $K$, $B$ and $D$ meson
systems \cite{Feyn.-rules,KB}. To simplify our discussion, we
consider two scenarios which can easily escape the constraints
\cite{LHT-2t,LHT-3t,eeppeq-LHT}:
\begin{eqnarray}
{\rm Case~I}:~~V_{H_d} &=& {\rm I},~
V_{H_u}=V^{\dag}_{\rm CKM}~~~~~~~~~~~~~~~~~~~~~~~~~~~~~~~~~~~~~~~~~~~~~~~~~~~~~~~~~~~~~~~~~~~~~\\
 {\rm Case~II}:~~s_{23}^{d}&=&1/{\sqrt{2}},~
s_{12}^{d}=s_{13}^{d}=0, ~\delta
_{12}^{d}=\delta_{23}^{d}=\delta_{13}^{d}=0~~~~~~~~~~~~~~~~~~~~~~~~~~~~~~
\end{eqnarray}
As for the mirror quark masses, it has been shown that the
experimental bounds on four-fermi interactions require $m_{H_i}\leq
4.8f^2/{\rm TeV}$\cite{EW constraint}. In our discussion, we take
this bound. We also assume a common mass for the first two
generation up-type mirror quarks, i.e. $m_{H_1}=m_{H_2}=500$ GeV and
let the third generation quark mass $m_{H_3}$ to vary from $600{\rm
GeV}$ to $1200{\rm GeV}$ for $f=500 {\rm GeV}$ and from $600{\rm
GeV}$ to $4800{\rm GeV}$ for $f=1000 {\rm GeV}$. To make our
predictions more realistic, we apply some kinematic cuts as did in
Ref. \cite{t-cggcut}, that is, we require the energy of each decay
product larger than 15 GeV in the top quark rest frame.

\begin{figure}
\setlength\subfigcapskip{-15pt} \vspace{-0.6cm}
\subfigure[][]{\includegraphics[width=3.0in,height=2.4in]{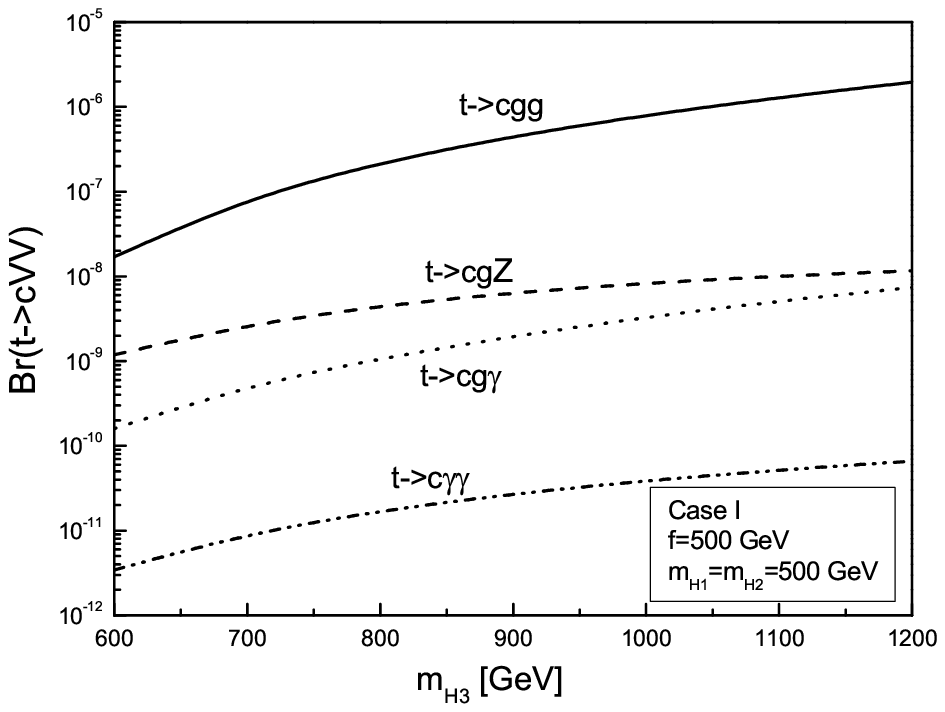}}
\hspace{-0.1in}%
\subfigure[][]{\includegraphics[width=3.0in,height=2.4in]{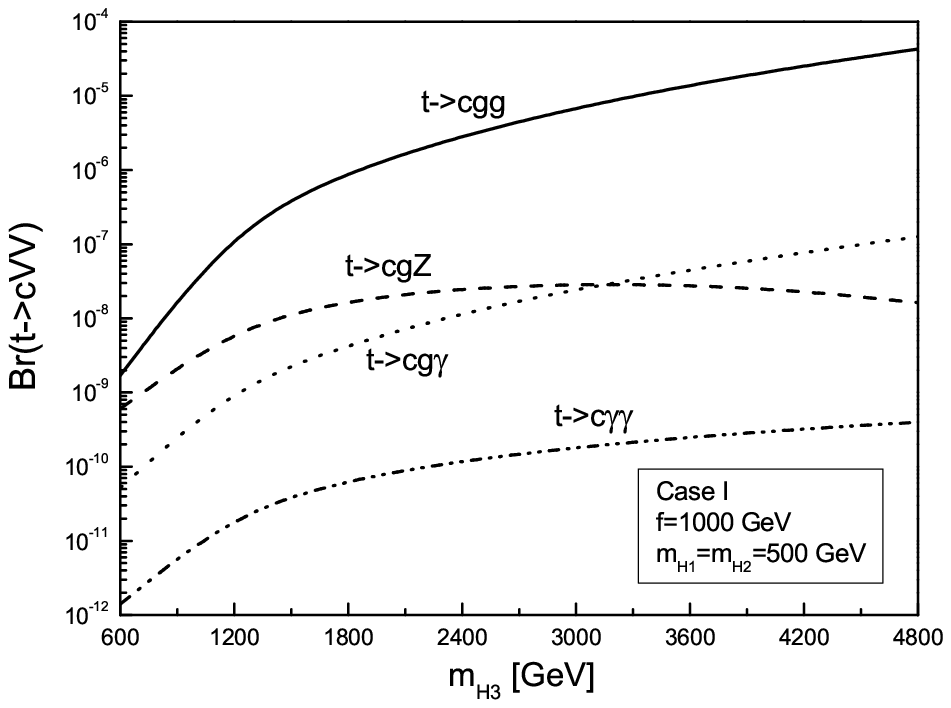}}
\hspace{-0.1in}%
\subfigure[][]{\includegraphics[width=3.0in,height=2.4in]{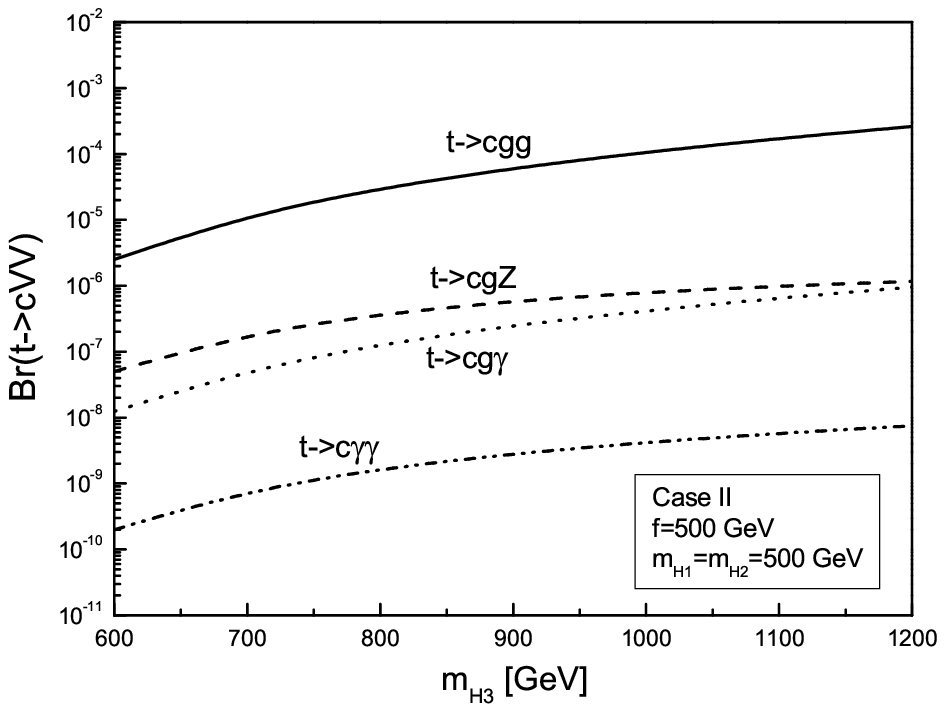}}
\hspace{-0.1in}%
\subfigure[][]{\includegraphics[width=3.0in,height=2.4in]{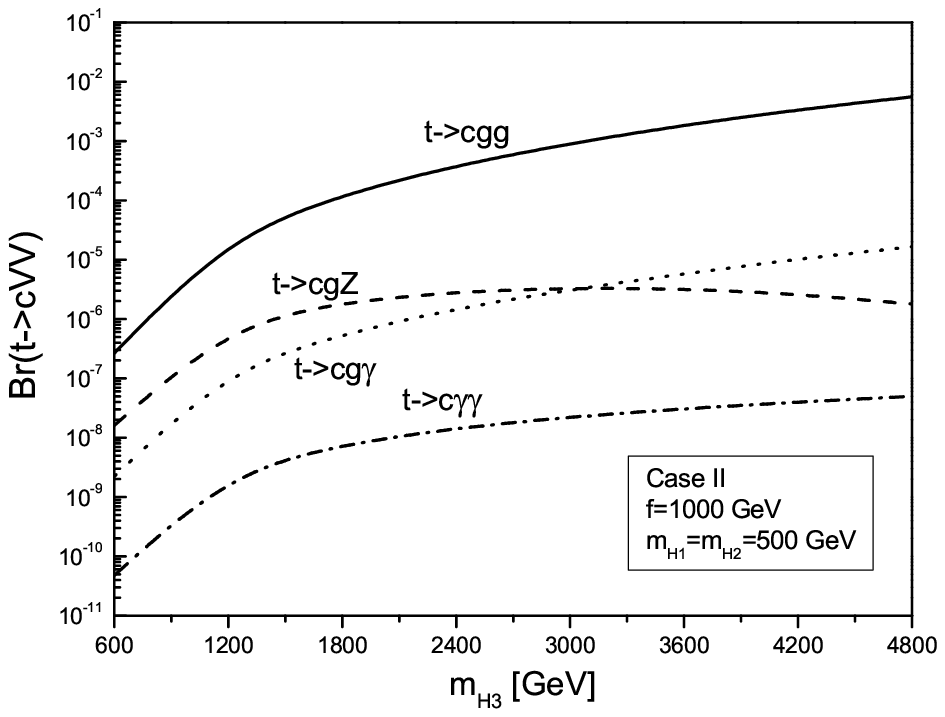}}
\vspace{0.01cm} \caption{\small The rates for
 $t\rightarrow cgg, cgZ, cg\gamma,c\gamma\gamma$ as a function of $m_{H_3}$
 for different values of $f$ and $V_{H_d}$.
 We take a common mass for the first two generation mirror quarks, i.e. $m_{H_1}=m_{H_2}=500$ GeV . \label{fig3}}
\end{figure}

\begin{figure}
\setlength\subfigcapskip{-15pt} \vspace{-0.0cm}
\subfigure[0.2][]{\includegraphics[width=3.0in,height=2.4in]{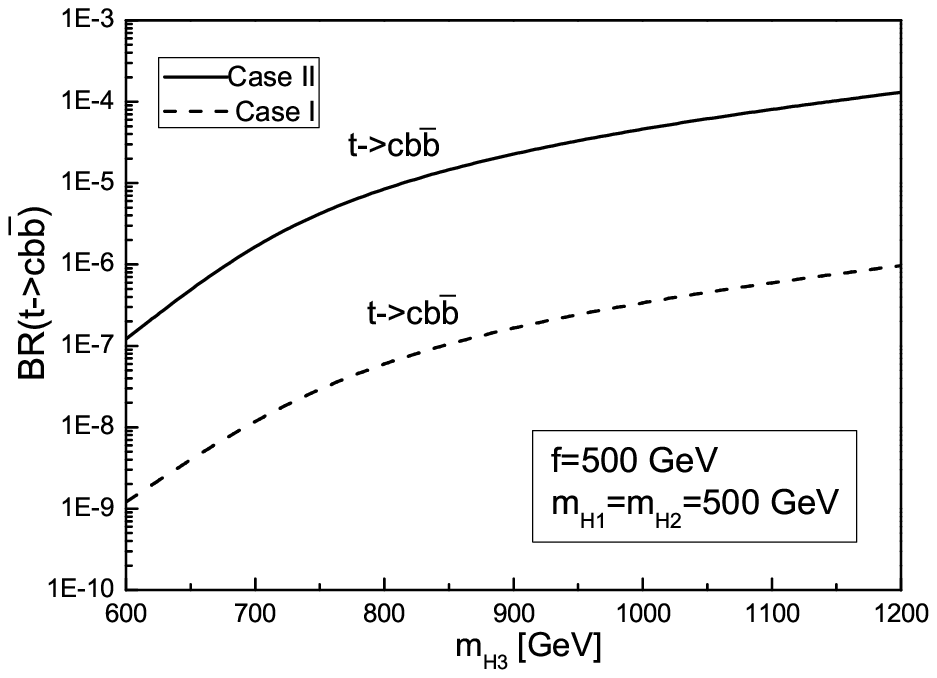}}
\hspace{-0.1in}%
\subfigure[][]{\includegraphics[width=3.0in,height=2.4in]{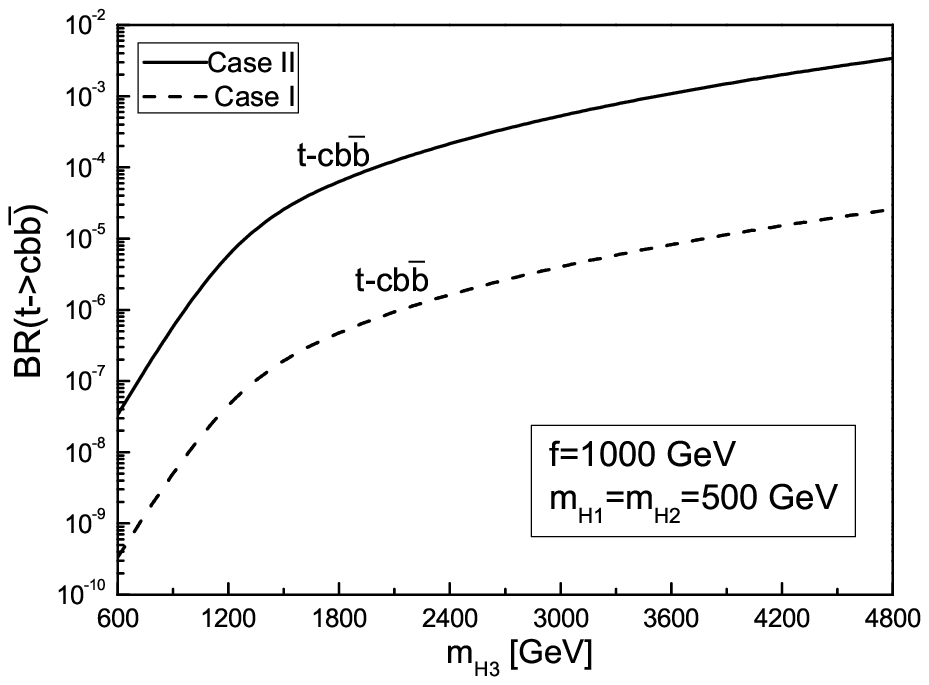}}
\hspace{-0.1in}%
\subfigure[][]{\includegraphics[width=3.0in,height=2.4in]{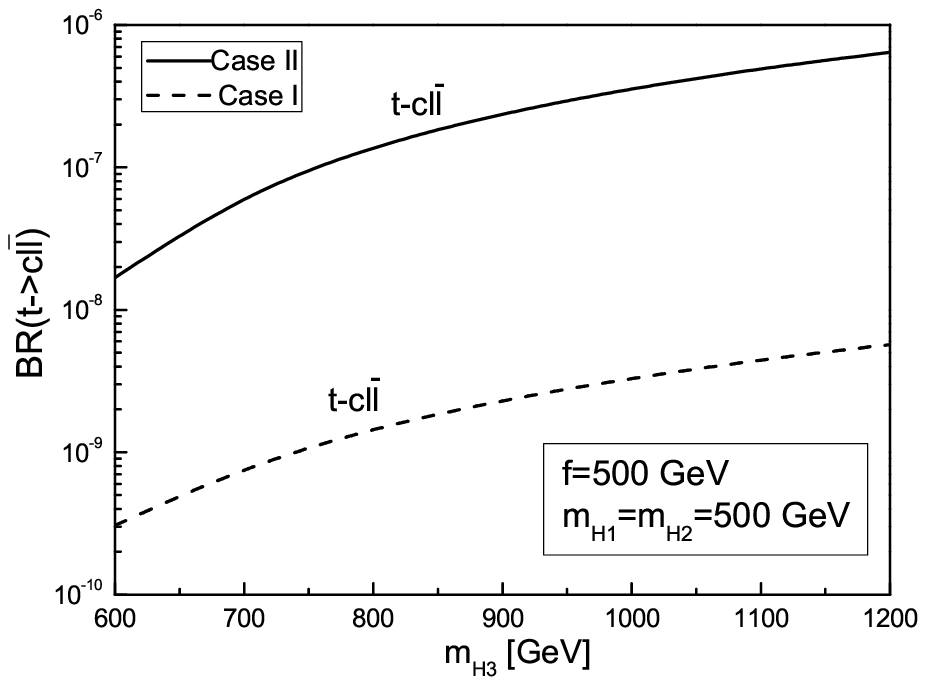}}
\hspace{-0.1in}%
\subfigure[][]{\includegraphics[width=3.0in,height=2.4in]{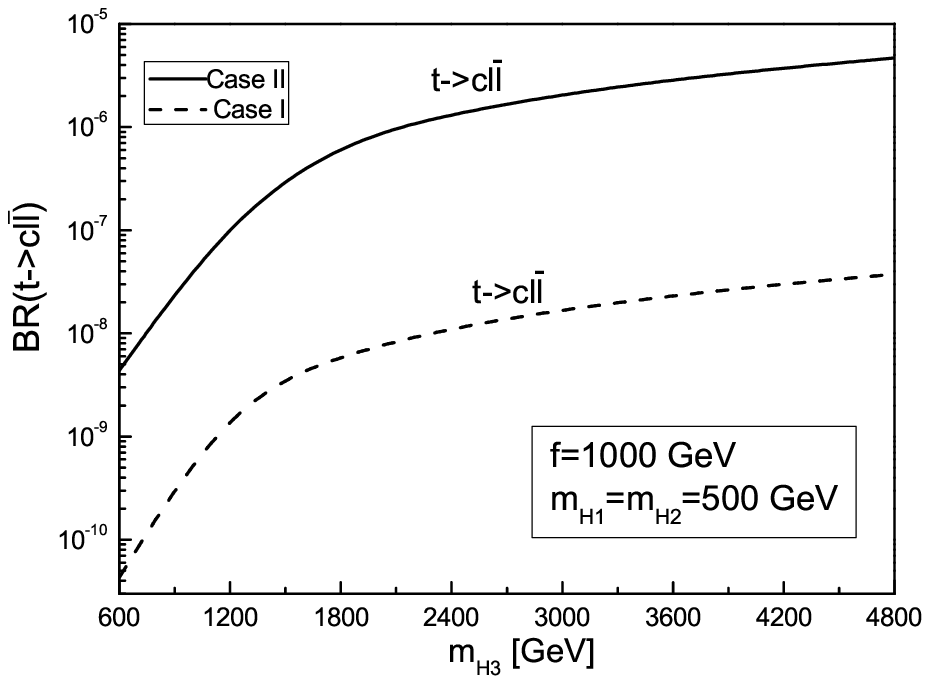}}
\vspace{0.01cm} \caption{\small Same as Fig.\ref{fig3}, but for the rates of
 $t\rightarrow cf\bar{f}~(f=b,e,\mu,\tau)$. }
\end{figure}

In Fig.3 we show the dependence of the rates for $t\rightarrow cgg,
cgZ, cg\gamma,c\gamma\gamma$ on $m_{H_3}$. This figure indicates
that the dependence is quite strong,  i.e. more than 1 order of
magnitude change when $m_{H_3}$ varies from 600 GeV to 1200 GeV in
Fig. (3a)and (3c) and from 600 GeV to 4800 GeV in Figs. (3b) and
(3d). The reason is, as explained in Eq. (\ref{formfactor}) and
below, the cancellation between the third generation mirror quark
contribution and the first two generation mirror quark contribution
is alleviated with the increase of $m_{H_3}$. This figure also
indicates that the rates $t\rightarrow cgg, cgZ,
cg\gamma,c\gamma\gamma$ are also sensitive to the parametrization
scenarios of $V_{H_d}$ when one compares the results in Fig.3 (a)
and Fig.3 (b) with those in  Fig.3 (c) and Fig.3 (d). This character
can be easily understood from the expression in Eq.
(\ref{formfactor}). From Fig.3, one may conclude that in the LHT
model, the rate for the decay $t\rightarrow cgg$ is much larger than
the others, reaching $10^{-3}$ in optimal cases, while the rates of
the decays $t\rightarrow cgZ, cg\gamma, c\gamma\gamma$ are all below
$10^{-5}$.

We investigate the same dependence of the decays $t\rightarrow
cf\bar{f}~(f=b,e,\mu,\tau)$ in Fig.4. Since the lepton masses are small compared with top quark mass,
 the rates for the decay $t\rightarrow c l\bar{l} $ with $l=e, \mu, \tau$ are approximately equal.
  This figure shows that that the
dependence of $t\rightarrow cf\bar{f}$ on $m_{H_3}$ is similar to
that of $t\rightarrow cVV$ shown in Fig.3. This figure also shows
the rate of the decay $t\rightarrow cb\bar{b}$  can reach $10^{-3}$
in the optimum case, while the rate of the decay $t\rightarrow
cl\bar{l}$  is usually less than $10^{-6}$.

The authors of \cite{collider} have roughly estimated the discovery potentials of the
high energy colliders in probing top quark FCNC decay for $100 fb^{-1}$ of integrated luminosity, and they obtained
\begin{eqnarray}
{\rm LHC}:Br(t\rightarrow cX)\geq5\times10^{-5}\\
{\rm ILC}:Br(t\rightarrow cX)\geq5\times10^{-4}\\
{\rm TEV33}:Br(t\rightarrow cX)\geq5\times10^{-3}
\end{eqnarray}
Then by the results presented in Figs. 3 and 4,  one can learn that
the LHT model can enhance the decays $t\rightarrow cgg(b\bar{b})$ to
the observable level of the LHC. So we may conclude that the LHC is
capable in testing the flavor structure of the LHT model.

\begin{center}
{\bf Table I: Optimum predictions for the decays $t\rightarrow cgg, cb\bar{b}, cl\bar{l}$ in different models.}\\
\doublerulesep 0.8pt \tabcolsep 0.008in
\begin{tabular}
{|c|c|c|c|c|c|c|}

\hline
     & SM & MSSM & TC2 & 2HDM & LHT Case I/Case II\\

\hline
    $Br(t\rightarrow cgg)$
   &$\mathcal {O}(10^{-9})$\cite{SM-3t}
   &$\mathcal{O}(10^{-4})$\cite{Eff.-vert.-MSSM-3t}
   &$\mathcal{O}(10^{-3})$\cite{TC2-tcgg}
   &$\mathcal {O}(10^{-3})$\cite{SM-2HDM-3t}
   &$\mathcal {O}(10^{-5})$ /$\mathcal {O}(10^{-3})$ \\

\hline
   $Br(t\rightarrow cl\bar{l})$
  &$10^{-14}$\cite{SM-MSSM-2HDM-3t}
  &$\mathcal{O}(10^{-7})$\cite{RPV-MSSM}
  &$\mathcal{O}(10^{-6})$\cite{tcll-TC2}
  &$\mathcal {O}(10^{-8})$\cite{2HDM-3t}
  &$\mathcal {O}(10^{-8})$ /$\mathcal {O}(10^{-6})$ \\

\hline
   $Br(t\rightarrow cb\bar{b})$
  &$\mathcal{O}(10^{-5})$\cite{tcbb-TC2}
  &$\mathcal{O}(10^{-7})$\cite{tcbb-TC2}
  &$\mathcal {O}(10^{-3})$\cite{tcbb-TC2}
  &---
  &$\mathcal {O}(10^{-5})$ /$\mathcal {O}(10^{-3})$ \\

\hline

\end{tabular}
\end{center}

Finally, we summarize the LHT model predictions for the FCNC
three-body decays $t\rightarrow cgg, cb\bar{b},cl\bar{l}$ in
comparison with the predictions of the SM, the MSSM, the TC2, and
the 2HDM in Table I. This table indicates that the optimum rates of
the decays in the LHT model are comparable with those in the TC2
model, and the predictions of the two models are significantly
larger than the corresponding predictions of the SM and the MSSM. As
far as the decay $Br (t\rightarrow cb\bar{b})$ is concerned, its
branching ratios may reach $10^{-3}$. So if the decays $t\rightarrow
cgg$ and $t\rightarrow cb\bar{b}$ are observed at the LHC,  more
careful theoretical analysis and  more precise measurement are
needed  to distinguish the models; while on the other hand, if these
decays are not observed, one can constrain the parameter space of
the LHT model. This table also indicates that, even in the optimum
cases, the rate for $t\rightarrow cl\bar{l}$ is only $10^{-6}$,
which implies that it is difficult to detect such decay.

\section{Conclusion}
In this work, we investigate the FCNC three-body decays
$t\rightarrow cVV ~(V=\gamma, Z, g)$ and $t\rightarrow
cf\bar{f}~(f=b, e, \mu, \tau)$ in the LHT. We conclude that: i) The
rates of these decays strongly depend on the mirror quark mass
splitting.  ii) The rates rely significantly on the flavor structure
of the mirror quarks, namely $V_{H_u}$ and $V_{H_d}$. iii) In the
optimum case of the LHT model, the rates for the decays
$t\rightarrow cgg$ and $t\rightarrow cb\bar{b}$  are large enough to
be observed at present or future colliders and  with the running of
the LHC, one get some useful information about the flavor structure
of the LHT model by detecting these decays.

\begin{acknowledgments} 

We would like to thank Junjie Cao and Lei Wu for helpful discussions
and suggestions. This work is supported by the National Natural
Science Foundation of China under Grant Nos.10775039, 11075045, by
Specialized Research Fund for the Doctoral Program of Higher
Education under Grant No.20094104110001, 20104104110001 and by
HASTIT under Grant No.2009HASTIT004.

\end{acknowledgments}

\newpage
{\large Appendix: Expressions of the effective $t\bar{c}V$ vertex}
\vspace{0.3cm}

The effective $t\bar{c}V$ vertex can be obtained by calculating
directly the diagrams in Fig.1 and with the help of the formula in
\cite{Eff.-vert.-MSSM-3t}. The loop functions in the effective
vertex are defined by the convention of \cite{BC-function} with
$p_t$ defined as the incoming momentum while  $p_c$ as the outgoing
momentum. In our calculation, higher order terms, namely, terms
proportional to $v^2/f^2$, in the masses of new gauge bosons and in
the Feynman rules are ignored.

\begin{eqnarray*}
&&\Gamma^{\mu}_{t\bar{c}\gamma}(p_c,p_t)
  = \Gamma^{\mu}_{t\bar{c}\gamma}(\eta)
   +\Gamma^{\mu}_{t\bar{c}\gamma}(\omega^{0})
   +\Gamma^{\mu}_{t\bar{c}\gamma}(\omega^{\pm})
   +\Gamma^{\mu}_{t\bar{c}\gamma}(A_{H})
   +\Gamma^{\mu}_{t\bar{c}\gamma}(Z_{H})
   +\Gamma^{\mu}_{t\bar{c}\gamma}(W_{H}^{\pm})\\
&&\quad \qquad\ \qquad\
   +\Gamma^{\mu}_{t\bar{c}\gamma}(W_{H}^{\pm}, \omega^{\pm}).\nonumber \\
&&\Gamma^{\mu}_{t\bar{c}\gamma}(\eta)
  = \frac{i}{16\pi^{2}}\frac{eg^{\prime2}}{150M_{A_{H}}^{2}}
   (V_{Hu})_{it}^*(V_{Hu})_{ic}(A+B+C), \\
&& A=\frac{1}{p_c^{2}-m_t^{2}}
     [ m_{Hi}^{2}(m_t^{2}B_{0}^{a}+p_c^{2}B_{1}^{a})\gamma^{\mu}P_{L}
      +m_tm_c(m_{Hi}^{2}B_{0}^{a}+p_c^{2}B_{1}^{a})\gamma^{\mu}P_{R}\\
&&\quad \quad
      +m_c(m_{Hi}^{2}B_{0}^{a}+m_t^{2}B_{1}^{a})\pslash_c\gamma^{\mu}P_{L}
      +m_{Hi}^{2}m_t(B_{0}^{a}+B_{1}^{a})\pslash_c\gamma^{\mu}P_{R}],\\
&& B=\frac{1}{p_t^{2}-m_{c}^{2}}
   [ m_{Hi}^{2}(m_c^{2}B_{0}^{b}+p_t^{2}B_{1}^{b})\gamma^{\mu}P_{L}
    +m_{t}m_c(m_{Hi}^{2}B_{0}^{b}+p_t^{2}B_{1}^{b})\gamma^{\mu}P_{R}\\
&&\quad \quad
    +m_cm_{Hi}^{2}(B_{0}^{b}+B_{1}^{b})\gamma^{\mu}\pslash_{t}P_{L}+m_t(m_{Hi}^{2}B_{0}^{b}
    +m_c^{2}B_{1}^{b})\gamma^{\mu}\pslash_{t}P_{R}], \\
&& C=[-m_{Hi}^{4}C_{0}^{1}\gamma^{\mu}P_{L}
      -m_{t}m_cm_{Hi}^{2}C_{0}^{1}\gamma^{\mu}P_{R}
      +m_{Hi}^{2}m_cC_{\alpha}^{1}\gamma^{\alpha}\gamma^{\mu}P_{L}\\
&&\quad\ \quad
     +m_cm_{Hi}^{2}(-\gamma^{\mu}\pslash_cC_{0}^{1}
     +\gamma^{\mu}\pslash_{t}C_{0}^{1}+\gamma^{\mu}\gamma^{\alpha}C_{\alpha}^{1})P_{L}
     +m_{Hi}^{2}m_tC_{\alpha}^{1}\gamma^{\alpha}\gamma^{\mu}P_{R} \\
&&\quad\quad\
     +m_{Hi}^{2}m_t(\gamma^{\mu}\gamma^{\alpha}C_{\alpha}^{1}
     -\gamma^{\mu}\pslash_cC_{0}^{1}+\gamma^{\mu}\pslash_{t}C_{0}^{1})P_{R}
     +m_{Hi}^{2}(\gamma^{\alpha}\gamma^{\mu}{\pslash_c}C_{\alpha}^{1}
     -\gamma^{\alpha}\gamma^{\mu}{\pslash_{t}}C_{\alpha}^{1} \\
&&\quad\quad
    -\gamma^{\alpha}\gamma^{\mu}\gamma^{\beta}C_{\alpha\beta}^{1})P_{L}
    +m_{t}m_c(\gamma^{\alpha}\gamma^{\mu}{\pslash_c}C_{\alpha}^{1}
    -\gamma^{\alpha}\gamma^{\mu}{\pslash_{t}}C_{\alpha}^{1}
    -\gamma^{\alpha}\gamma^{\mu}\gamma^{\beta}C_{\alpha\beta}^{1})P_{R}].\\
&& \Gamma^{\mu}_{t\bar{c}\gamma}(\omega^{0})
  = \frac{i}{16\pi^{2}}\frac{eg^{2}}{6M_{Z_{H}}^{2}}
   (V_{Hu})_{it}^*(V_{Hu})_{ic}(D+E+F),\\
&& D=A(B_{0}^{a}\rightarrow\ B_{0}^{c},B_{1}^{a}\rightarrow\ B_{1}^{c}),\\
&& E=B(B_{0}^{b}\rightarrow\ B_{0}^{d},B_{1}^{b}\rightarrow\ B_{1}^{d}),\\
&& F=C(C_{\alpha\beta}^{1}\rightarrow\ C_{\alpha\beta}^{2},
       C_{\alpha}^{1}\rightarrow\ C_{\alpha}^{2},
       C_{0}^{1}\rightarrow\ C_{0}^{2}).\\
&& \Gamma^{\mu}_{t\bar{c}\gamma}(\omega^{\pm})
  =\frac{i}{16\pi^{2}}\frac{eg^{2}}{2M_{W_{H}}^{2}}
   (V_{Hu})_{it}^*(V_{Hu})_{ic} [\frac{2}{3}G+\frac{2}{3}H-\frac{1}{3}I +J],\\
&&
 J= m_{Hi}^{2}m_{c}(p_{t}^{\mu}C_{0}^{4}-p_{c}^{\mu}C_{0}^{4}+2C_{\mu}^{4})P_{L}
   +m_{Hi}^{2}m_{t}(p_{t}^{\mu}C_{0}^{4}-p_{c}^{\mu}C_{0}^{4}+2C_{\mu}^{4})P_{R}\\
&&\quad \quad
   -m_{Hi}^{2}[p_{\mu\alpha}(p_{t}^{\mu}-p_{c}^{\mu}+2C_{\mu}^{4})+(p_{t}^{\mu}
   -p_{c}^{\mu})C_{\alpha}^{4}+2C_{{\mu}\alpha}^{4}]\gamma^{\alpha}P_{L}\\
&&\quad \quad
  -m_{t}m_c[p_{t\alpha}(p_{t}^{\mu}-p_{c}^{\mu}+2C_{\mu}^{4})+(p_{t}^{\mu}
  -p_{c}^{\mu})C_{\alpha}^{4}+2C_{{\mu}\alpha}^{4}]\gamma^{\alpha}P_{L},
\end{eqnarray*}
\begin{eqnarray*}
&& G=A(B_{0}^{a}\rightarrow\ B_{0}^{e},B_{1}^{a}\rightarrow\ B_{1}^{e}),\\
&& H=B(B_{0}^{b}\rightarrow\ B_{0}^{f},B_{1}^{b}\rightarrow\ B_{1}^{f}),\\
&& I=C(C_{\alpha\beta}^{1}\rightarrow\ C_{\alpha\beta}^{3},
       C_{\alpha}^{1}\rightarrow\ C_{\alpha}^{3},
       C_{0}^{1}\rightarrow\ C_{0}^{3}).\\
&& \Gamma^{\mu}_{t\bar{c}\gamma}(A_{H})
  =\frac{i}{16\pi^{2}}\frac{eg^{\prime2}}{75}(V_{Hu})_{it}^*(V_{Hu})_{ic}(K+L+M),\\
&& K=\frac{1}{p_{c}^{2}-m_{t}^{2}}[p_{c}^{2}
     B_{1}^{a}+m_{t}\pslash_cB_{1}^{a} ]\gamma^{\mu}P_{L},\\
&& L=\frac{1}{p_t^{2}-m_{c}^{2}}[p_{t}^{2}
     B_{1}^{b}+m_{c}\pslash_{t}B_{1}^{b} ]\gamma^{\mu}P_{L},\\
&&
M=[(\pslash_t-\pslash_{c})C_{\alpha}^{1}\gamma^{\mu}\gamma^{\alpha}
   -m_{Hi}^{2}C_{0}^{1}\gamma^{\mu}
   -C_{\alpha\beta}^{1}\gamma^{\alpha}\gamma^{\mu}\gamma^{\beta}]P_{L}.\\
&& \Gamma^{\mu}_{t\bar{c}\gamma}(Z_{H})
  =\frac{i}{16\pi^{2}}\frac{eg^{2}}{3}(V_{Hu})_{it}^*(V_{Hu})_{ic}(N+O+P),\\
&& N=K(B_{1}^{a}\rightarrow\ B_{1}^{c}),\\
&& O=L(B_{1}^{b}\rightarrow\ B_{1}^{d}),\\
&& P=M(C_{\alpha\beta}^{1}\rightarrow\ C_{\alpha\beta}^{2},
       C_{\alpha}^{1}\rightarrow\ C_{\alpha}^{2},
       C_{0}^{1}\rightarrow\ C_{0}^{2}).\\
&& \Gamma^{\mu}_{t\bar{c}\gamma}(W_{H}^{\pm})
  =\frac{i}{16\pi^{2}}\frac{eg^{2}}{2} (V_{Hu})_{it}^*(V_{Hu})_{ic}
   (\frac{4}{3}Q+\frac{4}{3}R-\frac{2}{3}S-T),\\
&&
T=[2\gamma^\mu(b_{0}+m_{WH}^{2}C_{0}^{4})+4\gamma^{\alpha}C_{\mu\alpha}+2(P^\mu_{t}
  -P^\mu_{c})\gamma^{\alpha}C_{\alpha}-\gamma^\mu\gamma^{\alpha}C_{\alpha}\\
&&\quad \quad
  +\gamma^\mu\pslash_t\gamma^{\alpha}C_{\alpha}+\gamma^\mu\gamma^{\alpha}(\pslash_c-\pslash_t)C_{\alpha}
  +2(\pslash_t-\pslash_c)\gamma^{\alpha}\gamma^\mu C_{\alpha}
  +\gamma^{\alpha}\pslash_t\gamma^\mu C_{\alpha}\\
&&\quad \quad
  +4\pslash_tC_\mu+2\pslash_t(p^\mu_t-p^\mu_c)+p^2_t\gamma^\mu C_{0}^{4}
  +\gamma^\mu\pslash_t\pslash_cC_{0}^{4}-2\pslash_t\pslash_c\gamma^\mu C_{0}^{4}]P_{L},\\
&& Q=K(B_{1}^{a}\rightarrow\ B_{1}^{e}),\\
&& R=L(B_{1}^{b}\rightarrow\ B_{1}^{f}),\\
&& S=M(C_{\alpha\beta}^{1}\rightarrow\ C_{\alpha\beta}^{3},
       C_{\alpha}^{1}\rightarrow\ C_{\alpha}^{3},
       C_{0}^{1}\rightarrow\ C_{0}^{3}).\\
&& \Gamma^{\mu}_{t\bar{c}\gamma}(W_{H}^{\pm}, \omega^{\pm})
  =\frac{i}{16\pi^{2}}\frac{g^{2}e}{2}(V_{Hu})_{it}^*(V_{Hu})_{ic}
   [m_t(\gamma^{\mu}\gamma^{\alpha}C_{\alpha}^{4}+2p^{\mu}_{t}C_{0}^{4}
     -\pslash_t\gamma^{\mu}C_{0}^{4})P_{R}\\
&&\quad\quad\ \ \ \ \quad\quad\ \ \ \ \
    +(m_c\pslash_tC_{0}^{4}+m_c\gamma^{\alpha}C_{\alpha}^{4}
    -2m_{Hi}^{2}C_{0}^{4})\gamma^{\mu}P_{L}].\\
&& \Gamma^{\mu}_{t\bar{c} Z}(p_c,p_t)
   = \Gamma^{\mu}_{t\bar{c}Z}(\eta)
    +\Gamma^{\mu}_{t\bar{c} Z}(\omega^{0})
    +\Gamma^{\mu}_{t\bar{c} Z}(\omega^{\pm})
    +\Gamma^{\mu}_{t\bar{c}Z}(A_{H})
    +\Gamma^{\mu}_{t\bar{c} Z}(Z_{H})
    +\Gamma^{\mu}_{t\bar{c}Z}(W_{H}^{\pm})\\
&&\quad \quad\quad \quad \ \ \quad
   +\Gamma^{\mu}_{t\bar{c}Z}(W_{H}^{\pm},\omega^{\pm}).\\
&& \Gamma^{\mu}_{t\bar{c}Z}(\eta)
  =\frac{i}{16\pi^{2}}\frac{g}{\cos\theta_{W}}
   \frac{g^{\prime2}}{100M_{A_{H}}^{2}}
   (V_{Hu})_{it}^*(V_{Hu})_{ic}(A'+B'+C'),\\
&& A'=\frac{1}{p_{c}^{2}-m_{t}^{2}}
     [(\frac{1}{2}-\frac{2}{3}\sin^{2}\theta_{W})
      m_{Hi}^{2}(m_t^{2}B_{0}^{a}+p_c^{2}B_{1}^{a})\gamma^{\mu}P_{L}
      -\frac{2}{3}\sin^{2}\theta_{W}m_tm_c(m_{Hi}^{2}B_{0}^{a}\\
&&\quad\quad
    +p_c^{2}B_{1}^{a})\gamma^{\mu}P_{R}+(\frac{1}{2}
    -\frac{2}{3}\sin^{2}\theta_{W})m_c(m_{Hi}^{2}B_{0}^{a}+m_t^{2}B_{1}^{a})\pslash_c\gamma^{\mu}P_{L}\\
&&\quad \quad
-\frac{2}{3}\sin^{2}\theta_{W}m_{Hi}^{2}m_t(B_{0}^{a}+B_{1}^{a})\pslash_c\gamma^{\mu}P_{R}],
\end{eqnarray*}
\begin{eqnarray*}
&& B'=\frac{1}{p_{t}^{2}-m_{c}^{2}}
     [(\frac{1}{2}-\frac{2}{3}\sin^{2}\theta_{W})m_{Hi}^{2}(m_c^{2}B_{0}^{b}
      +p_t^{2}B_{1}^{b})\gamma^{\mu}P_{L}
      -\frac{2}{3}\sin^{2}\theta_{W}m_{t}m_c(m_{Hi}^{2}B_{0}^{b}\\
&&\quad \quad
     +p_t^{2}B_{1}^{b})\gamma^{\mu}P_{R}+(\frac{1}{2}-\frac{2}{3}\sin^{2}\theta_{W})m_t(m_{Hi}^{2}B_{0}^{b}
     +m_c^{2}B_{1}^{b})\gamma^{\mu}\pslash_{t}P_{R}\\
&&\quad \quad
     -\frac{2}{3}\sin^{2}\theta_{W}m_cm_{Hi}^{2}(B_{0}^{b}+B_{1}^{b})\gamma^{\mu}\pslash_{t}P_{L}], \\
&&
  C'=(\frac{1}{2}-\frac{2}{3}\sin^{2}\theta_{W})C.\\
&& \Gamma^{\mu}_{t\bar{c}Z}(\omega^{0})
  =\frac{i}{16\pi^{2}}\frac{g}{\cos\theta_{W}}
   \frac{g^{2}}{4M_{Z_{H}}^{2}}(V_{Hu})_{it}^*(V_{Hu})_{ic}{m_{Hi}^{2}}(D'+E'+F'),\\
&& D'=A'(B_{0}^{a}\rightarrow\ B_{0}^{c},B_{1}^{a}\rightarrow\ B_{1}^{c}),\\
&& E'=B'(B_{0}^{b}\rightarrow\ B_{0}^{d},B_{1}^{b}\rightarrow\ B_{1}^{d}),\\
&& F'=C'(C_{\alpha\beta}^{1}\rightarrow\ C_{\alpha\beta}^{2},
         C_{\alpha}^{1}\rightarrow\ C_{\alpha}^{2},
         C_{0}^{1}\rightarrow\ C_{0}^{2}).\\
&& \Gamma^{\mu}_{t\bar{c}Z}(\omega^{\pm})
  =\frac{i}{16\pi^{2}}\frac{g}{\cos\theta_{W}}
   \frac{g^{2}}{2M_{W_{H}}^{2}}(V_{Hu})_{it}^*(V_{Hu})_{ic}(G'+H'+I'+J'),\\
&& G'=A'(B_{0}^{a}\rightarrow\ B_{0}^{e},B_{1}^{a}\rightarrow\ B_{1}^{e}),\\
&& H'=B'(B_{0}^{b}\rightarrow\ B_{0}^{f},B_{1}^{b}\rightarrow\ B_{1}^{f}),\\
&& I'=C'(C_{\alpha\beta}^{1}\rightarrow C_{\alpha\beta}^{3},
         C_{\alpha}^{1}\rightarrow\ C_{\alpha}^{3},
         C_{0}^{1}\rightarrow\ C_{0}^{3}),\\
&& J'=\cos^{2}\theta_WJ.\\
&& \Gamma^{\mu}_{t\bar{c}Z}(A_{H})
   =\frac{i}{16\pi^{2}}\frac{g}{\cos\theta_{W}}
    \frac{g^{\prime2}}{100}(V_{Hu})_{it}^*(V_{Hu})_{ic}(K'+L'+M'),\\
&&
K'=\frac{1}{p_{c}^{2}-m_{t}^{2}}
   [(\frac{1}{2}-\frac{2}{3}\sin^{2}\theta_{W})p_{c}^{2}B_{1}^{a}
    -\frac{2}{3}\sin^{2}\theta_{W}m_{t}\pslash_cB_{1}^{a}]\gamma^{\mu}P_{L},\\
&&
L'=\frac{1}{p_{t}^{2}-m_{c}^{2}}
  [(\frac{1}{2}-\frac{2}{3}\sin^{2}\theta_{W})p_{t}^{2}B_{1}^{b}
   -\frac{2}{3}\sin^{2}\theta_{W}m_{c}\pslash_{t}B_{1}^{b}]\gamma^{\mu}P_{L},\\
&&
M'=(\frac{1}{2}-\frac{2}{3}\sin^{2}\theta_{W})
   [(\pslash_t-\pslash_{c})C_{\alpha}^{1}\gamma^{\mu}\gamma^{\alpha}
   -m_{Hi}^{2}C_{0}^{1}\gamma^{\mu}
   -C_{\alpha\beta}^{1}\gamma^{\alpha}\gamma^{\mu}\gamma^{\beta}]P_{L}.\\
&& \Gamma^{\mu}_{t\bar{c}Z}(Z_{H})
  =\frac{i}{16\pi^{2}}\frac{g}{\cos\theta_{W}} \frac{g^{2}}{4}
  (V_{Hu})_{it}^*(V_{Hu})_{ic}(N'+O'+P'),\\
&& N'=K'(B_{1}^{a}\rightarrow\ B_{1}^{c}),\\
&& O'=L'(B_{1}^{b}\rightarrow\ B_{1}^{d}),\\
&& P'=M'(C_{\alpha\beta}^{1}\rightarrow\ C_{\alpha\beta}^{2},
         C_{\alpha}^{1}\rightarrow\ C_{\alpha}^{2},
         C_{0}^{1}\rightarrow\ C_{0}^{2}).\\
&& \Gamma^{\mu}_{t\bar{c} Z}(W_{H}^{\pm})
  =\frac{i}{16\pi^{2}}\frac{g}{\cos\theta_{W}}{g^{2}}
   (V_{Hu})_{it}^*(V_{Hu})_{ic}(Q'+R'+S'+T'),\\
&& Q'=K'(B_{1}^{a}\rightarrow\ B_{1}^{e}),\\
&& R'=L'(B_{1}^{b}\rightarrow\ B_{1}^{f}),\\
&& S'=(-\frac{1}{2}+\frac{1}{3}\sin^{2}\theta_{W})
     [(\pslash_t-\pslash_{c})C_{\alpha}^{3}\gamma^{\mu}\gamma^{\alpha}
     -m_{Hi}^{2}C_{0}^{3}\gamma^{\mu}-C_{\alpha\beta}^{3}\gamma^{\alpha}\gamma^{\mu}\gamma^{\beta}]P_{L},\\
&& T'=\cos^{2}\theta_{W}T.
 \end{eqnarray*}
\begin{eqnarray*}
&&\Gamma^{\mu}_{t\bar{c}Z}(W_{H}^{\pm}, \omega^{\pm})
  =\frac{i}{16\pi^{2}}\frac{g^{3}\cos\theta_W}{2}(V_{Hu})_{it}^*(V_{Hu})_{ic}
   [m_t(\gamma^{\mu}\gamma^{\alpha}C_{\alpha}^{4}+2p^{\mu}_{t}C_{0}^{4}
   -\pslash_t\gamma^{\mu}C_{0}^{4})P_{R}\\
&&\quad\quad\ \ \ \ \quad\quad\ \ \ \ \
   +(m_c\pslash_tC_{0}^{4}+m_c\gamma^{\alpha}C_{\alpha}^{4}-2m_{Hi}^{2}C_{0}^{4})\gamma^{\mu}P_{L}]\\
&& \Gamma^{\mu aij}_{t\bar{c}g}(p_c,p_t)
  = \Gamma^{\mu aij}_{t\bar{c}g}(\eta^{0})
   +\Gamma^{\mu aij}_{t\bar{c}g}(\omega^{0})
   +\Gamma^{\mu aij}_{t\bar{c}g}(\omega^{\pm})
   +\Gamma^{\mu aij}_{t\bar{c}g}(A_{H})
   +\Gamma^{\mu aij}_{t\bar{c}g}(Z_{H})
   +\Gamma^{\mu aij}_{t\bar{c}g}(W_{H}^{\pm}),\\
&& \Gamma^{\mu aij}_{t\bar{c}g}(\eta^{0})
  = \frac{i}{16\pi^{2}}\frac{g^{\prime2}}{100M_{A_{H}}^{2}}
    (V_{Hu})_{it}^*(V_{Hu})_{ic}g_sT^{aij}(A+B+C),\\
&& \Gamma^{\mu aij}_{t\bar{c}g}(\omega^{0})
  =\frac{i}{16\pi^{2}}\frac{g^{2}}{4M_{Z_{H}}^{2}}(V_{Hu})_{it}^*(V_{Hu})_{ic}g_sT^{aij}(D+E+F),\\
&& \Gamma^{\mu aij}_{t\bar{c}g}(\omega^{\pm})
  =\frac{i}{16\pi^{2}}\frac{g^{2}}{2M_{W_{H}}^{2}}(V_{Hu})_{it}^*(V_{Hu})_{ic}g_sT^{aij}(G+H+I),\\
&& \Gamma^{\mu aij}_{t\bar{c}g}(A_{H})
  =\frac{i}{16\pi^{2}}\frac{g^{\prime2}}{100}
  (V_{Hu})_{it}^*(V_{Hu})_{ic}g_sT^{aij}(K+L+M),\\
&& \Gamma^{\mu aij}_{t\bar{c}g}(Z_{H})
  =\frac{i}{16\pi^{2}}\frac{g^{2}}{4}(V_{Hu})_{it}^*(V_{Hu})_{ic}g_sT^{aij}(N+O+P),\\
&& \Gamma^{\mu aij}_{t\bar{c}g}(W_{H}^{\pm})
=\frac{i}{16\pi^{2}}\frac{g^{2}}{2}(V_{Hu})_{it}^*(V_{Hu})_{ic}g_sT^{aij}(Q+R+S).
\end{eqnarray*}
The two-point and three-point loop functions $B_0,B_1,C_0,~C_{ij}$
in the above expressions are defined as
\begin{eqnarray*}
&& C_{ij}^{1}=C_{ij}^{1}(-p_{c},p_{t},m_{Hi},M_{A_{H}},m_{Hi}),~
   C_{ij}^{2}=C_{ij}^{2}(-p_{c},p_{t},m_{Hi},M_{Z_{H}},m_{Hi}),\\
&& C_{ij}^{3}=C_{ij}^{3}(-p_{c},p_{t},m_{Hi},M_{W_{H}},m_{Hi}),~
   C_{ij}^{4}=C_{ij}^{4}(p_{c},-p_{t},M_{W_{H}},m_{Hi},M_{W_{H}}),\\
&& B^{a}=B^{a}(-p_{c},m_{Hi},M_{A_{H}}),~
   B^{b}=B^{b}(-p_{t},M_{Hi},M_{A_{H}}),\\
&& B^{c}=B^{c}(-p_{c},m_{Hi},M_{Z_{H}}),~
   B^{d}=B^{d}(-p_{t},M_{Hi},M_{Z_{H}}),\\
&& B^{e}=B^{e}(-p_{c},m_{Hi},M_{W_{H}}),~
   B^{f}=B^{f}(-p_{t},M_{Hi},M_{W_{H}}).
\end{eqnarray*}


\begin{thebibliography}{99}
\bibitem{1}
   J. A. Aguilar-Saavedra, Acta Phys. Polon. B {\bf35}, 2695 (2004).
\bibitem{sm-2t}
   G. Eilam, J. L. Hewett, and A. Soni, Phys. Rev. D {\bf44}, 1473 (1991);
                                                       {\bf59}, 039901 (1999);
   B. Mele, S. Petrarca, and A. Soddu, Phys. Lett. B {\bf435}, 401 (1998).

\bibitem{limit-LHC1}
   M. Beneke, I. Efthymipopulos, M. L. Mangano,  J. Womersley (conveners) {\em et al.}, arXiv: hep-ph/0003033.

\bibitem{limit-LHC2}
  J. Carvalho, N. Castro, A. Onofre, and F.Velosco (ATLAS
  Collaboration), ATLAS internal note, ATL-PHYS-PUB-2005-009, 2005.

\bibitem{limit-ILC}
  M. Cobal, AIP Conf. Proc. No. 753, (AIP, New York, 2005), p. 234.

\bibitem{detectable}
  W. Wagner, Rept. Prog. Phys.  {\bf68}, 2409 (2005);
  A. Juste {\em et~al.}, econf C0508141, PLEN0043 (2005);
  J. M. Yang, Ann. Phys. (N.Y.) {\bf316}, 529 (2005);
  D. Chakraborty, J. Konigsberg, and D. Rainwater, Ann. Rev. Nucl. Part. Sci. {\bf53}, 301 (2003);
  F. Larios, R. Martinez, and  M. A. Perez, Int. J. Mod. Phys. A {\bf21}, 3473 (2006).
\bibitem{MSSM-2t}
   C. S. Li, R. J. Oakes and J. M. Yang, Phys. Rev. D {\bf49}, 293 (1994);  {\bf56}, 3156 (1997);
   J. L. L\'{o}pez, D. V. Nanopoulos and R. Rangarajan, Phys. Rev. D {\bf56}, 3100 (1997);
   G. M. de Divitiis, R. Petronzio and L. Silvestrini, Nucl. Phys. B {\bf504}, 45 (1997);
   J. Guasch and J. Sola, Nucl. Phys. B {\bf562}, 3 (1999);
   J. J. Liu, C. S. Li, L. L. Yang and L. G. Jin, Phys. Lett. B {\bf599}, 92 (2004).
\bibitem{LR-SUSY}
   M. Frank and I. Turan, Phys. Rev. D {\bf72}, 035008 (2005).

\bibitem{SUSY-RV}
   J. M. Yang, B.-L.Young and X. Zhang, Phys. Rev. D {\bf 58}, 055001 (1998).


\bibitem{2HDM-2t}
   A. Arhrib, Phys. Rev. D {\bf72}, 075016 (2005);
   S. Bejar, J. Guasch, and J. Sola, Nucl. Phys. B {\bf675}, 270 (2003);
   E. O. Iltan, Phys. Rev. D {\bf65}, 075017 (2002);
   W. S. Hou, Phys. Lett. B {\bf296}, 179 (1992);
   R. A. D\'{i}az,  R. Mart\'{i}nez, and J. Alexis Rodr\'{i}guez, hep-ph/0103307.

\bibitem{TC2-2t}
   Xue-lei Wang, Gong-ru Lu, Jin-min Yang, {\em et al.}, Phys. Rev. D {\bf50}, 5781 ( 1994);
   Gong-ru Lu, Chong-xing Yue and Jin-shu Huang, J. Phys. G {\bf22}, 305 (1996);
                               Phys. Rev. D {\bf57}, 1755 (1998);
   Gong-ru Lu, Fu-rong Yin, Xue-lei Wang, and Ling-de Wan, Phys. Rev. D {\bf68}, 015002 (2003);
   Chong-xing Yue, Gong-ru Lu, Guo-li Liu, and Qing-jun Xu , Phys. Rev. D {\bf64}, 095004
   (2001).
\bibitem{Extra-quark}
   A. Arhrib and W. S. Hou, JHEP {\bf 0607}, (2006) 009.

\bibitem{SM-3t}
   E. Jenkins, Phys. Rev. D {\bf56}, 458 (1997);
   G. Altarelli, L. Conti and V. Lubicz, Phys. Lett. B {\bf502}, 125 (2001).

\bibitem{SM-3t-domin.}
   G. Eilam, M. Frank and I. Turan, Phys. Rev. D {\bf73}, 053011 (2006).
\bibitem{SM-2HDM-3t}
   S. Bar-Shalom, G. Eilam, M. Frank and I. Turan, Phys. Rev. D {\bf72}, 055018
   (2005);
   J. L. D\'{i}az-Cruz, M. A. P\'{e}rez, G. Tavares-Velasco, and J. J. Toscano, Phys. Rev. D {\bf60}, 115014 (1999).

\bibitem{SM-MSSM-2HDM-3t}
   M. Frank and I. Turan, Phys. Rev. D {\bf74}, 073014 (2006).
\bibitem{2HDM-3t}
   E. O. Iltan and I. Turan, Phys. Rev. D {\bf67}, 015004 (2003);
   S. Bar-Shalom, G. Eilam, A. Soni and J. Wudka, Phys. Rev. D {\bf57}, 2957 (1998).

\bibitem{MSSM-3t}
   G. Eilam, M. Frank and I. Turan, Phys. Rev. D {\bf74}, 035012
   (2006).

\bibitem{Eff.-vert.-MSSM-3t}
   J. J. Cao, G. Eilam, M. Frank, K. Hikasa, G. L. Liu, I. Turan, and J. M. Yang, Phys. Rev.
   D{\bf 75},075021(2007).
   \bibitem{RPV-MSSM}
   Zhaoxia Heng, Gongru Lu, Lei Wu, Jin Min Yang,  Phys. Rev. D {\bf 79}, 094029 (2009).


\bibitem{TC2-tcVV}
   Chong-xing Yue, Gong-ru Lu, Qing-jun Xu, {\em et al.}, Phys. Lett. B {\bf 508}, 290(2001).

\bibitem{TC2-tcgg}
   Huan-Jun Zhang,  Phys. Rev. D {\bf 77}, 057501 (2008).

\bibitem{tcll-TC2}
  Chong-xing Yue, Lei Wang, Dong-qi Yu Phys. Rev. D {\bf 70}, 054011 (2004);
  Chong-xing Yue, {\em et al.}, Mod. Phys. Lett. A {\bf 18}, 2187 (2003).

\bibitem{tcbb-TC2}
   Guoli Liu,  Chin. Phys. Lett. {\bf 26}, 101401(2009).

\bibitem{ind-3t}
   J. Drobnak, S. Fajfer and J. F. Kamenik,  JHEP, {\bf 0903}, 077 (2009).
\bibitem{LHT}
  H. C. Cheng and I. Low, JHEP, {\bf 0309}, 051 (2003);
  I. Low, JHEP, {\bf 0410}, 067 (2004);
  H. C. Cheng and I. Low,  JHEP, {\bf 0408}, 061 (2004);
  J. Hubisz and P. Meade,  Phys. Rev. D {\bf 71}, 035016 (2005).
\bibitem{LHT-2t}
   Hou Hong-Sheng, Phys. Rev. D {\bf 75}, 094010 (2007).

\bibitem{LHT-3t}
  Hui-Di Yang, Chong-Xing Yue, Jia Wen, Yong-Zhi Wang, Mod. Phys. Lett. A {\bf 24}, 1943 (2009).

\bibitem{LH1}
   N. Arkani-Hamed, A. G. Cohen, H. Georgi, Phys. Lett. B {\bf 513}, 232 (2001);
   N. Arkani-Hamed, A. G. Cohen, T. Gregoire, J. G. Wacker, JHEP {\bf 0208}, 020 (2002);
   N. Arkani-Hamed, A. G. Cohen, E. Katz, A. E. Nelson, T. Gregoire, J. G. Wacker, JHEP {\bf 0208}, 021 (2002).

\bibitem{LH2}
   N. Arkani-Hamed, A. G. Cohen, E. Katz, A. E. Nelson, JHEP {\bf 0207},  034 (2002)
  ;
   T. Han, H. E. Logan, B. McElrath and L. T. Wang, Phys. Rev. D {\bf 67}, 095004
   (2003).

\bibitem{Hierarchy}
   R. Barbieri, A. Strumia, Phys. Lett. B {\bf462}, 144 (1999).

\bibitem{Reintr.}
   W. Kilian, J. Reuter, Phys. Rev. D {\bf 70}, 015004 (2004);
   C. Csaki, J. Hubisz, G. D. Kribs, P. Meade, J. Terning, Phys. Rev. D {\bf67},  115002 (2003);
   J. L. Hewett, F. J. Petriello, T. G. Rizzo, JHEP {\bf 0310}, 062 (2003);
   C. Csaki, J. Hubisz, G. D. Kribs, P. Meade, J. Terning, Phys. Rev. D {\bf68}, 035009 (2003).




\bibitem{FCNC-LHT}
  M. Blanke, A. J. Buras, A. Poschenrieder, C. Tarantino, S. Uhlig, A. Weiler, JHEP {\bf 0612}, 003 (2006) ;
  J. Hubisz, S. J. Lee,  and G. Paz,  JHEP {\bf 0606}, 041 (2006).
  M. Blanke, A. J. Buras, S. Recksiegel, C. Tarantino, S. Uhlig, Phys. Lett. B {\bf657}, 81 (2007);
  M. Blanke, A. J. Buras, S. Recksiegel, C. Tarantino, S. Uhlig, JHEP {\bf 0706}, 082 (2007);
  M. Blanke, A. J. Buras, S. Recksiegel, C. Tarantino, arXiv:0805.4393v2 [hep-ph];
  M. Blanke, A. J. Buras, B. Duling, A. Poschenrieder, C. Tarantino, JHEP {\bf 0705}, 013 (2007).

\bibitem{Feyn.-rules}
 M. Blanke, A. J. Buras, A. Poschenrieder, S. Recksiegel, C. Tarantino, S. Uhlig, A. Weiler, JHEP {\bf 0701}, 066 (2007).
\bibitem{FC-LHT0}
   Andrzej J. Buras, Anton Poschenrieder, Selma Uhlig, and William A. Bardeen,
   JHEP {\bf 0611}, 062 (2006).



\bibitem{LoopTools}
   T. Hahn, M. Perez-Victoria, Comput. Phys. Commun. {\bf 118}, 153 (1999);
   T. Hahn, Nucl. Phys. Proc. Suppl. {\bf 135}, 333 (2004).


\bibitem{SM-paramet.}
   C. Amsler {\it et al}., Particle Data Group, Phys. Lett. B {\bf 667}, 1 (2008).

\bibitem{EW constraint}
   J. Hubisz, P. Meade, A. Noble, M. Perelstein,  JHEP {\bf 0601}, 135 (2006).

\bibitem {KB}
    M. Blanke etal.,  Phys. Lett.  B {\bf 646}, 253 (2007).

\bibitem{eeppeq-LHT}
   C. X. Yue, J. Wen, J. Y. Liu, W. Liu,  Chin. Phys. C {\bf 3}, 89 (2009);
   X. L. Wang, {\em et al.}, Nucl. Phys. B {\bf 807}, 210 (2009);
   X. L. Wang, {\em et al.},  Nucl. Phys. B {\bf 810}, 226 (2009).

\bibitem{t-cggcut}
   Jun jie Cao, Gad Eilam, Ken-ichi Hikasa, Jin Min Yang,  Phys. Rev. D {\bf 74}, 031701 (2006).

\bibitem {collider}
   J. Guasch and J. Sola,  Nucl. Phys. B {\bf 562}, 3 (1999).
\bibitem {BC-function}
  M. Clements, {\em et al.}, Phys. Rev. D {\bf 27}, 570 (1983);
  A. Axelrod, Nucl. Phys. B {\bf 209}, (1982) 349;
  G. Passarino, M. Veltman, Nucl. Phys. B {\bf 160} 151 (1979).

 \end{thebibliography}
\end{document}